\documentclass[conference]{IEEEtran}

\IEEEoverridecommandlockouts
\usepackage{cite}
\usepackage{amsmath,amssymb,amsfonts}
\usepackage{algorithmic}
\usepackage{graphicx}
\usepackage{textcomp}
\usepackage{xcolor}
\usepackage{booktabs}
\usepackage{textcomp}
\usepackage{xcolor}
\usepackage{multirow} % Required for the \multirow command
\usepackage{array} % For better column definitions if needed, though often not strictly required

\def\BibTeX{{\rm B\kern-.05em{\sc i\kern-.025em b}\kern-.08em
    T\kern-.1667em\lower.7ex\hbox{E}\kern-.125emX}}
\begin{document}

\title{Performance Evaluation of V2X Communication Using Large-Scale Traffic Data}

\author{\IEEEauthorblockN{John Pravin Arockiasamy}
\IEEEauthorblockA{
Karlsruhe Institute of Technology, Karlsruhe, Germany \\
john.arockiasamy@kit.edu}
\and
\IEEEauthorblockN{Alexey Vinel}
\IEEEauthorblockA{
Karlsruhe Institute of Technology, Karlsruhe, Germany \\
alexey.vinel@kit.edu}

}

\maketitle

\begin{abstract}
Vehicular communication (V2X) technologies are widely regarded as a cornerstone for cooperative and automated driving, yet their large-scale real-world deployment remains limited. As a result, understanding V2X performance under realistic, full-scale traffic conditions continues to be relevant. Most existing performance evaluations rely on synthetic traffic scenarios generated by simulators, which, while useful, may not fully capture the features of real-world traffic. In this paper, we present a large-scale, data-driven evaluation of V2X communication performance using real-world traffic datasets. Vehicle trajectories derived from the Highway Drone (HighD) and Intersection Drone (InD) datasets are converted into simulation-ready formats and coupled with a standardized V2X networking stack to enable message-level performance analysis for entire traffic populations comprising over hundred thousands vehicles across multiple locations. We evaluate key V2X performance indicators, including inter-generation gap, inter-packet gap, packet delivery ratio, and channel busy ratio, across both highway and urban intersection environments. Our results show that cooperative awareness services remain feasible at scale under realistic traffic conditions. In addition, the findings highlight how traffic density, mobility patterns, and communication range influence V2X performance and how synthetic traffic assumptions may overestimate channel congestion.
% Our results provide insights into the scalability and feasibility of cooperative awareness service under realistic deployment conditions and highlight the impact of traffic density, mobility patterns, and communication range on network performance.
\end{abstract}


\section{Introduction}

Despite several decades of intensive research, real-world deployment of vehicular communications (V2X) technologies remains limited~\cite{Uhlemann2024}. Although V2X has long been recognized as a key enabler of improved road safety -- allowing vehicles to exchange information beyond line of sight and to support cooperative driving -- large-scale adoption is still lacking. The most fundamental V2X building block, cooperative awareness~\cite{Lyamin2018}, has been implemented in only a small number of production vehicles. Consequently, understanding V2X performance under full-scale deployment conditions of vehicle-based V2X systems remains a subject of interest in the community. Numerous simulation-based studies have evaluated V2X networking performance~\cite{Masini2025,Wu2025}; however, they are typically based on synthetic traffic scenarios generated by traffic simulators, such as the widely used Luxembourg scenario in SUMO~\cite{codeca2015lust}. While valuable, such approaches are not always fully convincing to all stakeholders.

We argue for the use of large-scale real traffic datasets for V2X performance evaluation, in which vehicle communication behavior is analyzed on top of traffic behavior derived from real measurements while V2X communications are simulated over them. Moreover, our performance analysis is scalable, with computations performed for entire vehicle traffic populations (on the order of 100,000 vehicles) rather than being limited to a single or small fleet of ego vehicles. Our approach bridges two traditionally separate research communities: autonomous driving, which commonly relies on data-driven analysis~\cite{Stiller2022}, and V2X communications, which has largely focused on simulation-based evaluations~\cite{Rolich2025}. By coupling these two perspectives, we enable more realistic and scalable assessment of the V2X channel congestion as a key limiting networking factor for operation~\cite{DCC}.  

% Our contribution is twofold:
% \begin{itemize}
% \item We propose a V2X performance evaluation framework based on large-scale real-world traffic datasets, in which vehicle behavior is derived from measured trajectories while V2X communications are simulated, enabling more realistic assessment compared to purely synthetic scenarios.
% \item  We conduct scalable V2X performance analysis at the level of entire traffic populations -- on the order of hundred thousands of vehicles -- across both highway and urban intersection scenarios, providing quantitative insights into cooperative awareness performance and channel load under realistic traffic conditions.

This paper provides empirical evidence on large-scale V2X communication performance under realistic traffic conditions, rather than relying on synthetic or calibrated traffic demand. By deriving vehicle mobility data directly from measured real-world dataset, we enable population-scale evaluation of cooperative awareness without traffic model assumptions.

Our contributions are threefold:
\begin{itemize}
\item We present a data-driven V2X performance evaluation methodology that couples real-world traffic datasets with a standardized ETSI ITS-G5 networking stack, enabling message-level analysis under realistic highway and urban intersection traffic.
\item We conduct scalable V2X performance analysis at the level of entire traffic populations involving over 100,000 vehicles across multiple real-world locations, going beyond ego-centric or small-fleet studies.
\item We extract deployment-relevant insights on CAM generation behavior, distance-dependent reception performance, and channel load characteristics that cannot be observed using purely synthetic traffic scenarios.

\end{itemize}

The remainder of this paper is organized as follows. Section II describes the selection of real-world traffic datasets and methodology used to convert them into simulation-ready representations for V2X performance evaluation. Section III presents the evaluation metrics and discusses the V2X communication performance results obtained for both highway and urban intersection scenarios. Finally, Section IV concludes the paper and outlines directions for future work.

% Should be good if - Some reference related to IPG IGG Work from Others on Different Dataset. 

\section{Methods}

\subsection{Dataset Selection for V2X Performance Evaluation}

We evaluate the performance of vehicle-based V2X communication under realistic traffic conditions, making the selection of an appropriate dataset critical for obtaining meaningful insights. To this end, three key criteria are considered for dataset selection:  

(1) the dataset must be derived from \textit{real-world traffic data}, including detailed vehicle dynamics such as speed, position, and heading over time, thereby preserving realistic driver behavior, vehicle interactions, road geometry, and traffic heterogeneity. Synthetic or simulation-generated datasets are not sufficient, as they can introduce a domain gap and potentially lead to a simulation-reality discrepancy~\cite{suo2021trafficsim};
  
(2) the dataset should capture \textit{medium- to large-scale traffic scenarios}, with a substantial number of vehicles observed simultaneously and multiple vehicle classes (e.g., passenger cars, trucks) across diverse locations and road types, including highways and intersections, to realistically reflect traffic participants and network densities~\cite{zhou2025vision}; and  

(3) the dataset should cover a \textit{fixed roadway segment}, capturing all vehicles traversing that segment, thereby enabling analysis of multi-vehicle message exchange and collective awareness beyond an ego-vehicle-centric perspective, and supporting evaluation of the complex interactions inherent in real-world traffic~\cite{xu2023v2v4real}.

Some well-known simulation-based datasets, such as the Alicante–Murcia highway scenario~\cite{gonzalez2021alicante}, the Stuttgart Traffic scenario~\cite{forster2017evaluation}, the Luxembourg SUMO Traffic (LuST) scenario~\cite{codeca2015lust, codeca2017luxembourg}, the Ingolstadt Traffic Scenario (InTAS)~\cite{lobo2020intas}, and the Monaco SUMO Traffic (MoST) scenario~\cite{codeca2018monaco}, provide useful controlled environments but remain synthetic and therefore do not satisfy Criterion~(1). While these datasets approximate real traffic, they are not based on real traffic data and were thus not considered.  

Recent real-world datasets such as~\cite{xu2023v2v4real, kueppers2024v2aix, mei2022waymo} are collected from single ego-vehicle or small cooperative fleets, and remain inherently vehicle-centric. Multi-V2X~\cite{li2024multi}, although incorporating multiple connected automated vehicles and roadside units (RSUs), is produced through co-simulation and lacks raw real-world sensor recordings. Infrastructure-assisted datasets~\cite{xiang2024v2x, yu2023v2x, yu2022dair, zimmer2024tumtraf} provide real-world fixed-location coverage, but they do not fully capture all the vehicle dynamic and trajectory information in a segment across different locations and road types. As a result, these datasets do not fully satisfy all the criteria.  
%yu2023v2x, yu2022dair,

Among publicly available real-world multi-vehicle trajectory datasets, CN+~\cite{karunathilake2024cn+}, DeepScenario~\cite{dsc3d}, and the leveLX datasets~\cite{highDdataset, inDdataset} were evaluated against our criteria. CN+ provides large-scale naturalistic driving but is constrained to a single roadway geometry. DeepScenario and the leveLX datasets meet all three criteria, offering multiple locations with a large number of vehicles, detailed vehicle dynamics, diverse road layouts, and high-quality trajectories suitable for analyzing multi-vehicle V2X interactions.  

Considering dataset maturity, research community adoption, and coverage of diverse traffic scenarios across highways and urban intersections, the leveLX datasets, specifically the Highway Drone (HighD) Dataset~\cite{highDdataset} and the Intersection Drone (InD) Dataset~\cite{inDdataset}, were selected for this study.

\subsubsection{HighD Dataset}

The HighD dataset comprises real-world, top-down drone recordings of 60 tracks collected across six highway locations near Cologne, Germany. The recordings were captured in 4K at 25 frames per second (FPS) and cover approximately 420\,m of roadway, capturing 110{,}516 vehicles, including passenger cars, and trucks. The dataset represents large-scale vehicle traffic exhibiting naturalistic multi-lane highway behavior, making it well-suited for evaluating broadcast-based V2X communication in high-density, predominantly longitudinal traffic conditions. 

\subsubsection{InD Dataset}

The InD dataset consists of 33 tracks recorded at four urban intersections in Germany using the same top-down drone methodology. It covers approximately 8{,}233 vehicles, including cars and trucks, alongside 5{,}366 vulnerable road users (VRUs) such as pedestrians and bicycles, across intersection areas ranging from 80×40~m to 140×70~m. The dataset captures stop-and-go dynamics, yielding, and signalized control, introducing timing and interaction complexities not present in highway environments. These features make InD also suitable for evaluating vehicle-based V2X communication where intersection-specific mobility and communication delays are critical. While the dataset includes multiple VRUs, this work focuses mainly on vehicle data. VRUs are omitted from the present analysis but remain valuable for future studies on V2X interactions with non-motorized actors.

% \subsubsection{HighD Dataset}

% The HighD dataset comprises real-world, top-down drone recordings from 60 tracks/videos collected across six highway locations near Cologne, Germany. The recordings were captured in 4K at 25 frames per second (FPS) and cover approximately 420\,m of roadway with covering 110516 vehicles including passenger car, trucks, car with trailers etc. The dataset captures large-scale vehicles traffic exhibiting naturalistic multi-lane highway behavior, making it suitable for evaluating broadcast-based V2X communication in high-density, predominantly longitudinal traffic conditions.

% \subsubsection{InD Dataset}

% The InD dataset consists of 33 tracks/videos recorded at four real world urban intersections in Germany using the same top-down collection methodology which covers around 7887 vehicles such as cars, trucks and around 5366 vulnerable road users (VRUs) such as pedestrian, bicycle. Stop-and-go dynamics, signalized control, yielding, and conflict negotiation introduce timing and interaction complexities not present in highway environments, making InD suitable for evaluating V2X performance where communication delay influences intersection safety and throughput. Although the dataset contains multiple road-user types (vehicles, bicycles, pedestrians), this study focuses exclusively on vehicle trajectories, with other road-user types reserved for future work.

\begin{table}[ht]
\centering
\caption{Key V2X Communication Parameters Used}
\label{tab:v2x_parameters}
\begin{tabular}{l | l}
\toprule
\textbf{Parameter} & \textbf{Value / Description} \\ 
\midrule
V2X Stack & ETSI ITS-G5 \\
Simulator Integration & OMNeT++ / Artery / SUMO \\
Carrier Frequency & 5.9~GHz (DSRC) \\
Channel Number & 180 \\
Channel Bandwidth & 10~MHz  \\
Transmission Power & 200~mW \\
Receiver Energy Detection & -85~dBm \\
Receiver Sensitivity & -82 dBm \\
CAM Size  & 85, 285 bytes \\
CAM Generation Interval & 100--1000~ms (configurable)\\
MAC Protocol & IEEE 802.11p / EDCA \\
Decentralized Congestion Control (DCC) & Reactive DCC \\
\bottomrule
\end{tabular}
\end{table}

\subsection{Experimental Setup}

The HighD and InD datasets do not natively include V2X communication data between vehicles. However, these datasets can be leveraged to evaluate V2X performance by treating all vehicles as V2X-capable nodes that generate Cooperative Awareness Messages (CAMs)~\cite{Lyamin2018} within a multi-agent communication environment. This approach enables the study of large-scale vehicle communication, even though real-world vehicles in the recordings were not equipped with V2X technology. 

To facilitate the V2X capability, the datasets must be reproduced in the \textit{SUMO}~\cite{sumo} microscopic traffic simulator and coupled with the \textit{Artery}~\cite{riebl2015artery} V2X networking stack. Artery, built on \textit{OMNeT++}\footnote{https://omnetpp.org/} and integrating the \textit{Vanetza}\footnote{https://www.vanetza.org/} ETSI ITS-G5 protocol stack, allows all vehicles to be treated as V2X-capable nodes generating, transmitting, and receiving CAM. The key communication parameters used in Artery, are summarized in Table~\ref{tab:v2x_parameters}, enabling synchronized evaluation of vehicle mobility and message-level performance.

However, neither HighD nor InD provides SUMO compliant formats. Therefore, their original raw data must be converted into the SUMO compliant formats. This conversion is a critical first step to ensure accurate, repeatable, and large-scale V2X communication performance evaluation while preserving the temporal and spatial fidelity of the original recordings.

\subsection{Dataset Conversion}

The PEGASUS project formalized a five-layer ontology for describing traffic scenarios on German highways~\cite{bagschik2018ontology}. Street-level geometry (L1), traffic infrastructure (L2), and temporal modifications to these static components (L3) constitute the \textit{static scenario elements}. In contrast, \textit{dynamic scenario elements} include movable road users (L4) and environmental conditions (L5) which influence vehicle behavior and sensing performance. Collectively, these layers provide the semantic foundation required to represent and simulate real-world traffic environments. Both the HighD and InD datasets can be interpreted using this layered traffic scenario concept.

\subsubsection{Conversion of Static Scenario Elements}

The HighD and InD datasets define static elements using \textit{Lanelet2}~\cite{poggenhans2018lanelet2} and ASAM \textit{OpenDRIVE}\footnote{https://www.asam.net/standards/detail/opendrive/}
 formats. However, the SUMO microscopic simulation framework requires static elements to be represented in the \textit{SUMO Network XML (.net.xml)} format. To bridge this incompatibility, SUMO \textit{netconvert}\footnote{https://sumo.dlr.de/docs/netconvert.html}, an official and standard SUMO tool, is used to convert OpenDRIVE format into SUMO Network format. This conversion produces a topologically accurate road representation with nodes and junctions, preserving lane connectivity, and intersection layouts for subsequent simulation stages.

\subsubsection{Conversion of Dynamic Scenario Elements}

Both HighD and InD deliver detailed dynamic elements that include: per-frame $(x,y)$ positions, vehicle classification, physical dimensions, velocities, accelerations, heading directions, and more, encapsulated within three core CSV files per track: \textit{track, trackMeta, and recordingMeta}. 

% A Python visualization framework is provided with the datasets to enable inspection and validation of trajectories prior to simulation.

For execution within SUMO, dynamic elements must be represented as \textit{SUMO Route XML (.rou.xml)} file. Since neither HighD nor InD provides vehicle routes compatible with SUMO, their CSV data must be programmatically converted into SUMO route files. While tools such as SUMO \textit{TraCI}\footnote{https://sumo.dlr.de/docs/TraCI.html} can use CSV data directly into SUMO, this method is incompatible with the Artery V2X networking stack, which is essential for evaluating V2X communication performance. Therefore, each vehicle dynamics must be reconstructed on a frame-by-frame basis to ensure that the original patterns are faithfully preserved within the SUMO simulation environment.

Let the vehicle position data in the original datasets be denoted by $(x_{\mathrm{Origveh}}, y_{\mathrm{Orig veh}})$, expressed relative to a location-specific origin $(x_{\mathrm{Orig origin}}, y_{\mathrm{Orig origin}})$. The converted SUMO road network, generated using the SUMO \textit{netconvert} tool, adopts its own coordinate origin $(x_{\mathrm{SUMOorigin}}, y_{\mathrm{SUMOorigin}})$. To align vehicle positions with the SUMO network across all track CSV files, a coordinate transformation is applied to map vehicle positions from the original dataset to the SUMO network as follows:

\begin{equation}
\begin{split}
x_{\mathrm{SUMOveh}} &= x_{\mathrm{SUMOorigin}} - (x_{\mathrm{Origorigin}} - x_{\mathrm{Origveh}}), \\
y_{\mathrm{SUMOveh}} &= y_{\mathrm{SUMOorigin}} - (y_{\mathrm{Origorigin}} - y_{\mathrm{Orig veh}}).
\end{split}
\label{eq:coord_transform}
\end{equation}

\noindent
SUMO route reconstruction is performed by iteratively associating each transformed vehicle position with the closest node or junction in the SUMO network. For a vehicle position ($x_{\mathrm{SUMOveh}}, y_{\mathrm{SUMOveh}}$) in SUMO and a set of network nodes or junctions indexed by $i \in \{1,2,...,N\}$ with coordinates $(x_i, y_i)$, the euclidean distance to all nodes in meters is computed as:

\begin{equation}
d_i = \sqrt{(x_{\mathrm{SUMOveh}} - x_i)^2 + y_{\mathrm{SUMOveh}} - y_i)^2}, \quad \forall i \in N.
\label{eq:euclid_multi}
\end{equation}

\noindent
The index of the closest node or junction is determined as:

\begin{equation}
\hat{i} = \arg \min_{i \in N} (d_i),
\label{eq:argmin}
\end{equation}

\noindent
The assignment of node or junction is performed only if the minimum distance satisfies a predefined threshold  $d_{\max}$:

\begin{equation}
d_{\hat{i}} \leq d_{\max}.
\label{eq:threshold}
\end{equation}

\noindent
If this condition is not met, the current frame is skipped, and the algorithm proceeds to the next frame, as one frame within a node or junction is enough to associate the vehicle position to its corresponding. Even if a vehicle temporarily fails to satisfy the threshold, the route can be reconstructed continuously by interpolating between previously and subsequently associated nodes, ensuring trajectory continuity across all frames. 

A grid-search algorithm was implemented to determine an optimal $d_{\max}$ in the range 0--5\,m with 0.5\,m increments, balancing spatial accuracy and robust node or junction association across diverse road geometries. A threshold of $d_{\max}=4$\,m was selected as it achieved this balance across all track CSV files. Sequential evaluation of all frames produces an ordered list of nodes or junctions traversed by the vehicles, forming its reconstructed routes. 

Based on these reconstructed routes, SUMO-compliant \texttt{.rou.xml} route files are generated per track CSV file for each vehicle at 25~FPS, matching the sampling rate of the original recordings. All relevant dynamic parameters required for SUMO, including acceleration, deceleration, maximum speed, vehicle class, departure position, departure speed, and arrival speed, are directly extracted from the CSV files and mapped to the reconstructed routes. This preserves the temporal and behavioral fidelity of the original dataset, ensuring that motion dynamics, inter-vehicle spacing, and stop-and-go patterns are represented while remaining compatible with SUMO's traffic rules. Although the dataset conversion pipeline is a necessary technical enabler, the primary contribution of this work lies in the empirical insights obtained from the evaluation of the performance of V2X under real-world large-scale mobility patterns, rather than in the conversion process itself.

To guaranty the integrity of the conversion process, a subsequent post-dataset conversion quality assessment procedure is performed, confirming that the generated SUMO routes faithfully reproduce the original dataset and are suitable for large-scale V2X performance evaluation.

\subsection{Post-Dataset Conversion Analysis}

After converting the original \textit{HighD} and \textit{InD} datasets into SUMO-compatible formats, we conducted a location-wise analysis by grouping all vehicle tracks according to their respective recording locations, covering all six HighD and four InD locations. A location-specific evaluation is essential, as traffic dynamics and V2X communication performance can vary significantly between locations, even within the same type. This approach enables a more accurate and detailed assessment of the fidelity of the dataset transformation.

To evaluate the fidelity of the conversion, two primary metrics are employed:

\begin{enumerate}
    \item \textbf{Number of Vehicles preserved}, which verifies that all vehicles present in the original datasets are retained in the converted SUMO scenarios without loss.
    \item \textbf{Average running time difference}, which compares the mean running time of vehicles between the original and SUMO datasets.
\end{enumerate}

The \textit{average running time difference} metric is selected because it implicitly aggregates multiple aspects of vehicle behavior, including speed, acceleration and deceleration patterns, stop durations, and interactions with traffic. By verifying that the average running time remains consistent, we ensure that the converted SUMO scenarios preserve the overall temporal dynamics of real traffic.

The \textit{average running time difference} ($\Delta t$) is computed as the average across all vehicles for each location and is defined as
\begin{equation}
\Delta t_{\ell} = \frac{1}{N_{\ell}} \sum_{i=1}^{N_{\ell}} \left( T^{\text{SUMO}}_{i,\ell} - T^{\text{Original}}_{i,\ell} \right),
\end{equation}
where $\ell$ denotes the location index, $N_{\ell}$ is the total number of vehicles observed at location $\ell$, and $T^{\text{SUMO}}_{i,\ell}$ and $T^{\text{Original}}_{i,\ell}$ represent the running time of vehicle $i$ at location $\ell$ in the SUMO-converted and original datasets, respectively.

\subsubsection{HighD}

After converting the original HighD dataset to SUMO format, we assess the transformation across all six recording locations. The 60 tracks were grouped according to their respective locations, and the evaluation was performed..

Table~\ref{highd_conversion} summarizes the number of vehicles and the corresponding $\Delta t$ for each location in both the original data and the SUMO data. The results show that the number of vehicles is preserved exactly for all locations, with a total of 110,516 vehicles in both the original and SUMO-converted datasets. This confirms that the conversion pipeline does not introduce any vehicle loss.

\begin{table}[ht]
\centering
\caption{Conversion Summary for HighD Dataset}
\label{highd_conversion}
\begin{tabular}{c |c| c |c}
\toprule
\textbf{Location} & \textbf{Vehicles (Orig.) } & \textbf{Vehicles (SUMO)} & \boldmath$\Delta t$ \textbf{ [s]} \\ 
\midrule
1 & 85962 & 85962  & 0.400 \\ 
2 & 3074  & 3074  & 0.696 \\ 
3 & 3747  & 3747   & 0.336 \\ 
4 & 4751  & 4751   & 0.669 \\ 
5 & 10079 & 10079  & 0.686 \\ 
6 & 2903  & 2903   & 1.217 \\ 
% \midrule
% \multicolumn{3}{r|}{\textbf{Total Vehicles}} & \textbf{--} & \textbf{--} & \textbf{110{,}516} \\
% \multicolumn{3}{r|}{\textbf{Average Difference}} & \textbf{--} & \textbf{--} & \textbf{0.667} \\
\bottomrule
\end{tabular}
\end{table}

The average running time difference of vehicles exhibits small but consistent differences between the original and SUMO datasets. Location~1, despite having the highest vehicle count, shows only a 0.4~s running time difference per vehicle. Location~6 presents the largest deviation (1.217~s), while Location~3 shows the smallest difference (0.336~s).

These small differences are expected. SUMO strictly enforces traffic rules and collision-free behavior, resulting in slightly more conservative motion compared with human driving in the original HighD recordings. Since HighD represents multi-lane highway traffic, even minor variations in acceleration or lane-change timing of leading vehicles can propagate downstream and manifest as small timing differences in the simulation.

% Overall, the variations remain minor relative to the total trip durations, and the SUMO-converted dataset preserves the structure, vehicle counts, and traffic scale of the original HighD dataset.

% \begin{table}[ht]
% \centering
% \caption{Conversion Summary for HighD Dataset}
% \label{highd_conversion}
% \begin{tabular}{c c c c c c}
% \hline
% Location &
% Vehicles (Orig.) &
% Vehicles (SUMO) &
% Avg. Orig (s) &
% Avg. SUMO (s) &
% $\Delta t$ (s) \\
% \hline
% 1 & 85962 & 85962 & 14.534 & 14.934 & 0.400 \\
% 2 & 3074  & 3074  & 13.358 & 14.054 & 0.696 \\
% 3 & 3747  & 3747  & 13.296 & 13.632 & 0.336 \\
% 4 & 4751  & 4751  & 12.036 & 12.705 & 0.669 \\
% 5 & 10079 & 10079 & 13.381 & 14.067 & 0.686 \\
% 6 & 2903  & 2903  & 13.134 & 14.351 & 1.217 \\
% \hline
% \end{tabular}
% \end{table}

\subsubsection{InD}

The InD dataset captures vehicle trajectories at urban intersections. As shown in Table~\ref{ind_conversion}, all vehicle trajectories from 33 tracks across the four locations were successfully grouped according to their respective locations and converted into SUMO-compatible formats without any loss, preserving a total of 8{,}233 vehicles. 
% It is important to note that InD Location~1 includes some tracks as small construction-site in Location 1, however, no major statistically meaningful differences were observed. Consequently, all Location~1 tracks are treated as a single, including the construction-site data, for the remainder of this paper. 

It is important to note that InD Location~1 includes some tracks recorded under temporary construction-site conditions. However, no statistically significant differences were observed in the analysis. Consequently, all tracks from InD Location~1, including those from the construction-site scenario, are treated as a single location for the remainder of this paper.

\begin{table}[ht]
\centering
\caption{Conversion Summary for InD Dataset}
\label{ind_conversion}
\begin{tabular}{c |c| c| c|c}
\toprule
\textbf{Location} & \textbf{Veh (Orig.) } & \textbf{Veh (SUMO)} & \textbf{Parked Veh.} & \boldmath$\Delta t$ \textbf{ [s]} \\ 
\midrule
 1 & 2503 & 2503 & 63 & -0.055 \\
2 & 2436 & 2436 & 228 & 0.859 \\
3 & 1196 & 1196 & 28 & 4.191 \\
4 & 2098 & 2098 & 27 & 2.300 \\ 
% \midrule
% \multicolumn{3}{r|}{\textbf{Total Vehicles}} & \textbf{--} & \textbf{--} & \textbf{110{,}516} \\
% \multicolumn{3}{r|}{\textbf{Average Difference}} & \textbf{--} & \textbf{--} & \textbf{0.667} \\
\bottomrule
\end{tabular}
\end{table}

\begin{figure*}[t]
    \centering
    \includegraphics[width=0.95\textwidth]{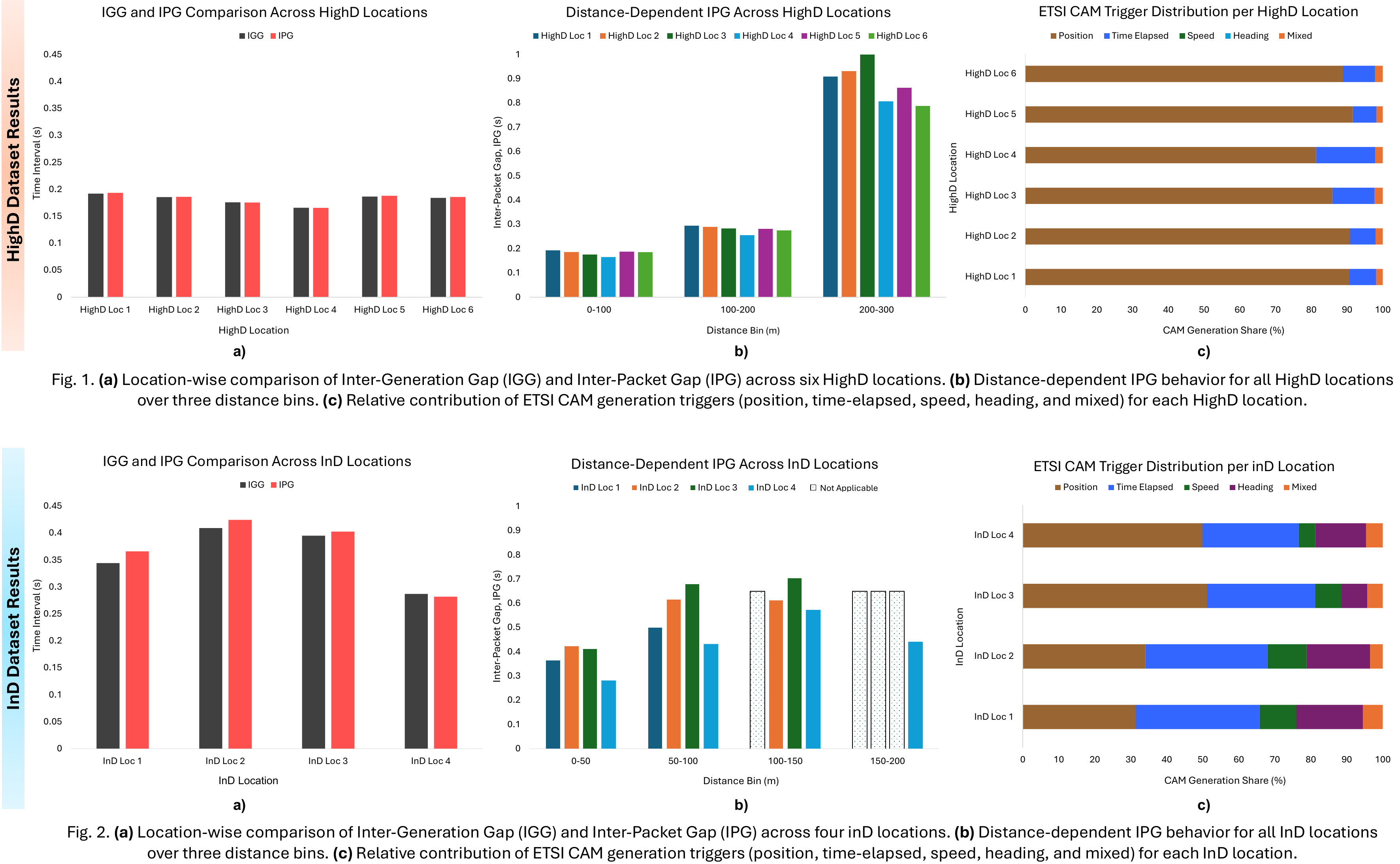}
    % \caption{IGG, IPG, and PDR across six HighD locations.}
    \label{fig:highD_distancewise_ipg}
\end{figure*}

The average running time difference in all vehicles are more noticeable than in the HighD dataset. The largest positive deviation occurs at Location~3 (4.191~s), while Location~1 exhibits a slight negative difference ($-0.055~\mathrm{s}$), indicating that the vehicles in SUMO simulation is marginally faster than the original recording. These variations are expected, as complex intersection dynamics and stop-and-go behavior amplify minor discrepancies between real data and SUMO’s rule-based traffic. Vehicles may move slightly more cautiously or differently, resulting in longer or shorter running times in some cases.

Additionally, Table~\ref{ind_conversion} shows the number of parked vehicles in each location. For the analyzes that follow, parked vehicles are not included, as they are inactive and do not participate in CAM transmission or reception.

% \begin{table}[ht]
% \centering
% \caption{Conversion Summary for InD Dataset}
% \label{ind_conversion}
% \begin{tabular}{c c c c c c}
% \hline
% Location &
% Vehicles (Orig.) &
% Vehicles (SUMO) &
% Avg. Orig (s) &
% Avg. SUMO (s) &
% $\Delta t$ (s) \\
% \hline
% 1 & 2503 & 2503 & 9.480 & 9.424 & -0.055 \\
% 2 & 2436 & 2436 & 14.850 & 15.709 & 0.859 \\
% 3 & 1196 & 1196 & 8.741 & 12.932 & 4.191 \\
% 4 & 2098 & 2098 & 10.522 & 12.822 & 2.300 \\
% \hline
% \end{tabular}
% \end{table}

\section{Results and Discussion}

\subsection{Key Performance Indicator (KPI) Definitions}

In this study, we use common and well-known V2X performance evaluation metrics to analyze the full deployment evaluation of V2X using the HighD and InD datasets on a per-location basis, enabling analysis of location-specific performance variations.. The metrics used for this analysis are as follows: 

\textbf{Inter-Generation Gap (IGG):}  
The average time interval between consecutively generated V2X messages from the same vehicle throughout the simulation time.

\textbf{Inter-Packet Gap (IPG):}  
The average time interval between consecutively received packets from a reference (transmitting) vehicle at a receiving vehicle over the duration of the vehicle’s simulation. In this study, IPG is evaluated additionally as a distance-dependent metric within a specific range for both datasets.

\textbf{Packet Delivery Ratio (PDR):}  
The ratio of number of successfully received packets at a receiver to number of transmitted packets from a reference transmitter. PDR is a fundamental metric for evaluating V2X communication performance~\cite{dettinger2024survey}. 

\textbf{Generation Cause:}  
A trigger condition responsible for generating a CAM as defined in~\cite{ETSI-CAM}, associated with thresholds such as a 1\,s time elapsed duration, a speed change of 0.5\,m/s, a heading change of 4°, a position change of 4\,m, or a mixed trigger driven by multiple simultaneous changes.

\textbf{Channel Busy Ratio (CBR):}  
The percentage of channel time sensed as busy due to ongoing transmissions. Additionally, we use \textit{mean CBR} which is defined as the channel utilization averaged across all vehicles within each individual location.
% In this work, the \textit{mean CBR} indicates the channel utilization averaged across all the locations.
%~\cite{correa2019infrastructure}

\subsection{KPIs Evaluation: HighD Dataset}

The HighD dataset comprises vehicle trajectories characterized by uninterrupted multilane traffic and predominantly longitudinal motion, and the V2X KPI results for this dataset across different locations are illustrated in Fig~1. 

The observed IGG for CAMs across all vehicles and locations ranges from 0.166~s to 0.192~s, which is consistent with the ETSI-defined CAM generation interval. This behavior is expected due to continuous vehicle motion and the absence of frequent stops or abrupt maneuvers typically observed in highway traffic.

\begin{figure*}[t]
    \centering
    \includegraphics[width=0.9999\textwidth]{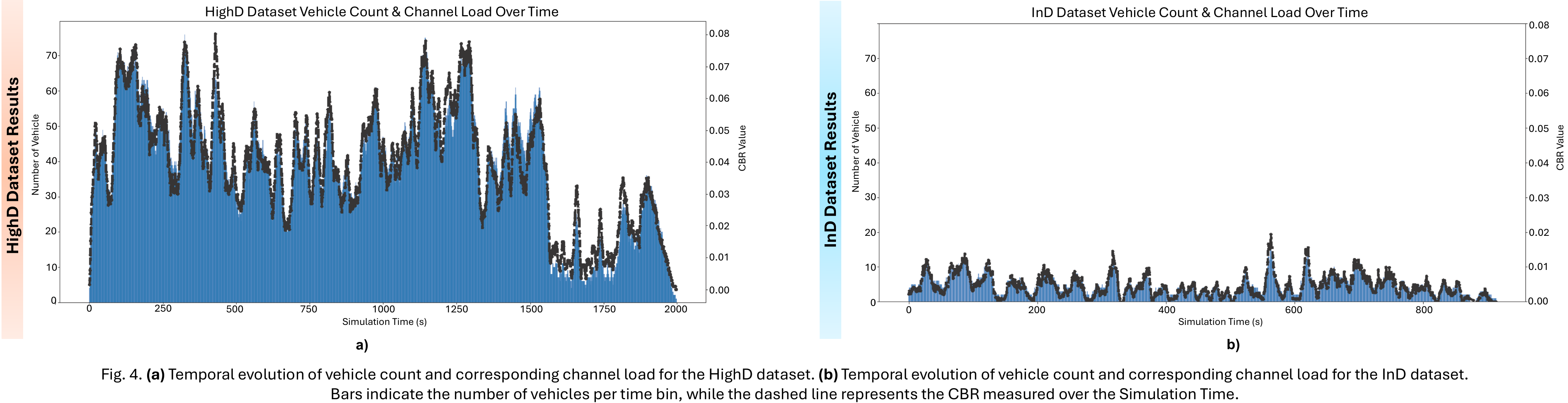}
    % \caption{IGG, IPG, and PDR across four InD locations.}
    \label{fig:ind_distancewise_ipg}
\end{figure*}

CAM generation is dominated by position-based triggers, which account for more than 80\% of CAMs at all locations. Time-elapsed conditions contribute a smaller but consistent share (above 7\%), while speed- and mixed-trigger conditions contribute only marginally. Heading-based triggers are not observed, indicating limited directional variation.

The IPG closely follows the IGG at all locations, ranging from 0.166~s to 0.193~s across all the vehicles. However, a distance-segmented IPG analysis reveals a clear degradation trend with increasing communication range. Vehicles within 0--100~m maintain stable IPG values similar to the overall average, while packets received at distances of 100--200~m experience noticeable delays. Beyond 200~m, the IPG values increase significantly to 0.807--1.014~s, indicating delayed reception of successive CAMs. These results show that, under realistic highway traffic conditions, communication range clearly limits the cooperative awareness performance. Despite large vehicle populations and continuous CAM generation, reception degradation occurs mainly at longer distances.
Finally, the PDR remains close to 100\% in all locations, confirming that the number of messages received matches the number of expected transmissions.

\subsection{KPIs Evaluation: InD Dataset}

The detailed evaluation of KPIs on the InD dataset across different locations is shown in Fig~2. The InD dataset captures vehicle behavior, where traffic signals, stop-and-go motion, and multi-directional trajectories introduce higher mobility variability. This is reflected in increased IGG values compared to HighD, ranging from approximately 0.287~s to 0.409~s. These higher IGG values reflect the intermittent nature of intersection traffic, where vehicles frequently stop, accelerate, and change direction.

CAM generation triggers in the InD dataset are more evenly distributed than in HighD scenarios. Across all locations, position-based triggers account for 31--51\% of CAM generation, while time-elapsed triggers contribute between 26--34\%. Heading based triggers, which are absent in HighD, contribute between 7--18\%, and speed based triggers contribute up to 10\%, highlighting the impact of turning maneuvers and frequent trajectory changes typical of urban intersections.

The IPG in InD shows slightly higher variability than in HighD, reflecting increased communication overhead and mobility uncertainty. Distance-segmented analysis indicates that IPG values remain close to the overall average for distances below 50~m, increase to 0.432--0.680~s for distances between 50~m and 150~m, and become inreliable beyond 150~m, with measurable data available only for Location~4 with less vehicles. This behavior indicates that effective cooperative awareness range in intersection environments is significantly shorter than on highways, due to rapid topology changes and limited line-of-sight conditions. The resulting PDR ranges from approximately 94.3\% to 100\% across different locations, indicating higher reception variability compared to highway environments.

\subsection{Channel Load Performance: HighD and InD Dataset}

We evaluated channel load performance on each dataset across all locations, illustrated in Fig~3. On average, the HighD dataset exhibits a higher channel load performance than the InD dataset, primarily due to the substantially larger number of active vehicles. Nevertheless, even in the HighD dataset, the observed \textit{mean CBR} values remain well below 0.1, indicating that the channel load is well within acceptable limits\cite{etsi2021cbr} when considering realistic vehicle densities in real-world scenarios. These results indicate that realistic traffic density alone is insufficient to induce critical channel congestion, even under full V2X penetration assumptions.

%, with overall \textit{mean CBR} values of 0.002 for InD and 0.017 for HighD across all locations.

\begin{figure}[t]
    \centering
    \includegraphics[width=0.45\textwidth]{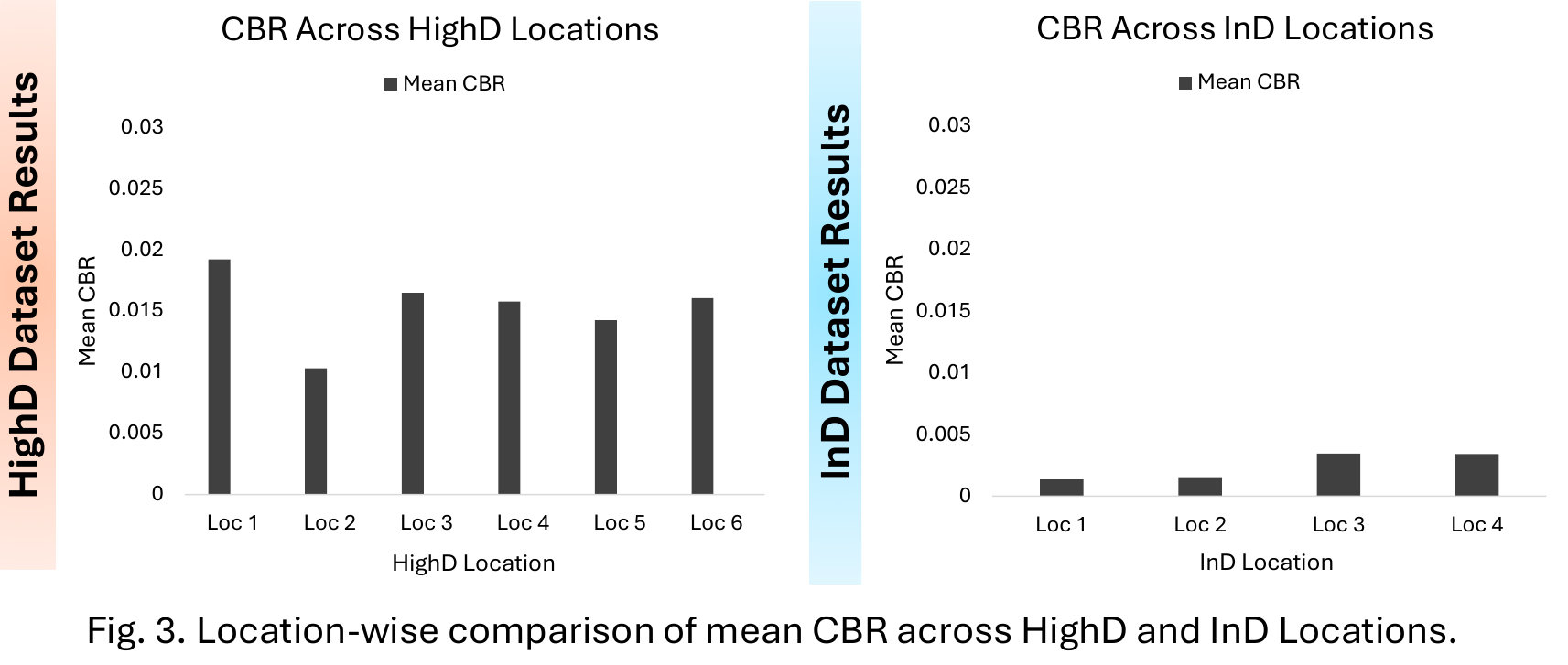}
    % \caption{IGG, IPG, and PDR across four InD locations.}
    \label{fig:CBR_distancewise_ipg}
\end{figure}

While average CBR values provide a high-level overview of channel utilization, they do not fully capture worst-case or localized congestion scenarios. To capture such effects, we analyze a single track with the highest CBR contribution from each dataset and examine its evolution over simulation time, vehicle count, and CBR value, as illustrated in Fig~4.

For the HighD dataset, Track~25 is selected as it exhibits the highest CBR contribution. As illustrated in Fig.~4(a), this track spans nearly 2000~s of SUMO simulation time and reaches a peak CBR of approximately 0.080 when the vehicle count is around 78. In contrast, Track~2 is selected from the InD dataset. As shown in Fig.~4(b), a pronounced CBR peak is observed between 500~s and 600~s, coinciding with a maximum vehicle count of 16 and resulting in a CBR value of approximately 0.020. Even in these worst-case tracks, CBR peaks remain short-lived and closely correlated with brief increases in local vehicle density. No sustained channel congestion is observed over time. This trend confirms that CBR increases only when vehicle density rises locally, and drops quickly when traffic disperses, indicating that congestion is temporary rather than sustained.

% This trend indicates a clear correlation between vehicle density and channel load: an increase in the number of vehicles leads to a higher volume of CAM transmissions, thereby increasing the CBR, while a reduction in vehicle count results in lower channel occupancy. 

Overall, realistic traffic dynamics inherently limit prolonged channel saturation, suggesting that channel congestion control mechanisms in real deployments may be triggered less frequently than predicted by simulation studies.

\subsection{Impact of Reactive DCC on CAM suppression: HighD and InD Dataset}

To investigate the impact of reactive DCC on both transmission and reception-side CAM suppression across all locations in both datasets, ETSI reactive DCC was enabled in the Artery network stack with the configurations mentioned in Table~\ref{tab:v2x_parameters}, allowing both transmission and reception-side suppression mechanisms to operate automatically.

\begin{table}[ht]
\centering
\caption{Mean DCC Transmission and Reception Suppression Ratios Across Datasets}
\label{tab:dcc_suppression}
\begin{tabular}{c | c | c}
\toprule
\textbf{Dataset} & \textbf{Mean Tx Suppression} & \textbf{Mean Rx Suppression} \\ 
\midrule
HighD & 0.000 & 0.000 \\
InD & $\approx$0.000 & 3.098\% \\
\bottomrule
\end{tabular}
\end{table}

Table~\ref{tab:dcc_suppression} summarizes the mean transmission and reception suppression ratios observed across all locations for each dataset. For the HighD dataset, both transmission-side and reception-side suppression remain zero, consistent with the earlier observation that channel load does not exceed the thresholds required to trigger ETSI reactive DCC mechanisms. 

In contrast, the InD dataset exhibits a non-zero mean reception-side suppression ratio, while transmission-side suppression remains negligible. Although the \textit{mean CBR} for the InD dataset remains below the DCC activation thresholds, reception-side suppression is observed due to localized bursts of overlapping CAMs and receiver processing limitations, allowing receivers to temporarily discard packets under transient load conditions. This behavior indicates that, in urban intersection scenarios, reactive DCC primarily responds to localized reception-side congestion rather than transmission overload.

\section{Conclusion and Future Work}

This paper presented a scalable, data-driven evaluation of V2X communication performance using large-scale real-world traffic datasets reproduced within a simulation-based V2X environment. By integrating HighD and InD dataset with SUMO and the Artery V2X networking stack, the study enables communication-level analysis under realistic highway and urban intersection traffic conditions. The results demonstrate that real deployment of cooperative awareness services are feasible at scale, while clearly illustrating the influence of traffic density, mobility patterns, and roadway context on V2X performance metrics. The analysis further shows that ETSI reactive DCC is rarely activated under realistic traffic and only, suggesting that real-world deployments may experience significantly fewer congestion control interventions.

Future work will extend this framework to evaluate additional V2X services, with a particular focus on Collective Perception~\cite{CO2026}, and to incorporate interactions involving non-motorized road users, especially vulnerable road users~\cite{VRU2025, lobo2025vru}. In addition, we aim to refine and expand the set of performance metrics to enable a more comprehensive assessment of V2X communication. These extensions will further support realistic, data-driven evaluation of V2X systems under complex and heterogeneous traffic scenarios.

\section*{Acknowledgment}
This research is funded by SIMON "Safe mobility and navigation through predictive risk management using swarm intelligence and V2X communication" mFund project (2024-2027). The authors gratefully acknowledge levelxdata\footnote{https://levelxdata.com/} for providing access to the HighD dataset and InD Dataset. These datasets were used exclusively for academic research purposes within the scope of this work.

\bibliographystyle{IEEEtran}
\bibliography{reference}\textbf{}
\end{document}